\documentclass[prd,aps,nofootinbib]{revtex4}
\usepackage{amssymb}
\usepackage{graphicx}
\usepackage{amsmath}
\usepackage{epstopdf}

\begin{document}

\title{\textbf{On the Casimir energy for a massive quantum scalar field and the Cosmological constant}}

\date{\today}

\author{G. Gazzola$^{(a)}$}\email[]{ggazzola@fisica.ufmg.br}
\author{M. C. Nemes$^{(a)}$}\email[]{carolina@fisica.ufmg.br}
\author{W. F. Wreszinski$^{(b)}$}\email[]{wreszins@fma.if.usp.br}

\affiliation{(a) Universidade Federal de Minas Gerais - Departamento de F\'{i}sica - ICEx \\
P.O. BOX 702, 30.161-970, Belo Horizonte MG - Brazil}
\affiliation{(b) Universidade de S\~{a}o Paulo - Departamento de F\'{i}sica Matem\'{a}tica - Instituto de F\'{i}sica \\
P.O. BOX 66318, 05.315-970, S\~{a}o Paulo SP - Brazil}

\vspace{5.0cm}

\begin{abstract}
We present a rigorous, regularization independent local quantum field theoretic treatment of the Casimir effect for a quantum scalar field of mass $\mu\ne0$ which yields closed form expressions for the energy density and pressure. As an application we show that there exist  special states of the quantum field  in which the expectation value of the renormalized energy-momentum tensor is, for any fixed time, independent of the space coordinate and  of the perfect fluid form 
$g_{\mu,\nu}\rho$ with $\rho>0$, thus providing a concrete quantum field theoretic model of the Cosmological constant. This $\rho$ represents the energy density associated to a state consisting of the vacuum and a certain number of excitations of zero momentum, i.e., the constituents correspond to lowest energy and pressure $p\le0$.
\end{abstract}

\maketitle
\newpage
\section{Introduction and Summary}

In a recent very interesting paper, Dappiaggi, Fredenhagen and Pinamonti [1] studied stable cosmological models driven by a free quantum scalar field, under the hypothesis that the underlying algebraic quantum state is of Hadamard type [2], and showed that, in the massive case, and under certain reasonable assumptions on the solutions of the semiclassical Einstein equations, an effective cosmological constant emerges. In order to achieve this, the authors address the important topic of the most general form of the renormalised stress-energy tensor associated to meaningful semiclassical solutions of the Einstein's equations.

In this paper we look at the problem from the point of view of the local quantum field theory (lqft) of the Casimir effect introduced in [3] and developped in [4] and [5]. As proposed in [3], measuring the (renormalised) Hamiltonian density 
$H(x)$ (see eq. (7) below) is a local operation: in order to measure $H(x)$, our measuring apparatus needs only examine a small neighborhood ,say $N$, of $x$, whose observable algebra has precisely the same structure as if it were embedded in globally flat space-time, and thus needs no information from global topology.  What does depend on the global structure is the state of the system. There exists, however, a proviso: how small may $N$ be? It cannot be reduced to the point $x$, because of the local structure of quantum fields, but may, in priciple, be arbitrarily small: on the other hand, the energy density is not bounded below (as an operator valued distribution) as $N$ contracts to a point [6]. This large indeterminacy, further discussed below, is the main subject of our paper. We work at fixed time (which may be thought as the cosmic time in applications to general relativity) and absorb the indeterminacy involved in the choice of $N$ in the dimension of a ``small'' region of fixed shape - we take it for simplicity to be a cube of side $L$- to which we ``restrict'' a free scalar quantum field by imposing some boundary condition (b.c.). One might expect that the state of the system be independent of this ``local'' b.c., but that is not the case. Indeed, as proved in [5], only \textbf{periodic} b.c. are compatible with lqft for a \textbf{scalar} quantum field and, clearly, there is a global change of spatial topology in passing to periodic b.c.. The word ``local'' means, in this special context, that
$$
L \ll L_c
\eqno{(1a)}
$$

where $L_c$ denotes a characteristic length of the system. In Cosmology, $L_c=L_H$, where $L_H$ is the Hubble length (see, e.g., [13]). we shall come back to (1a) in section IV.

A special issue has acquired special interest in present-day Cosmology: are there low-energy states of a scalar quantum field with the properties:
a.) negative pressure;
b.) the cosmological constant condition
$$ 
p=-\rho
\eqno{(1b)}
$$
where $p$ is the pressure and $\rho$ the energy density corresponding to the given state. From a.) and b.) there results
c.) 
$$\rho\ge0
\eqno{(1c)}$$

The precise significance of (1a)-(1c) and the relation to the cosmological constant is discussed in section IV. We point out, however, that condition a.) is independent of the cosmological constant relation b.) and is a well-known property of dark energy, a model of which is considered as an application of the results of section III and part of section IV at the end of the latter section (after (36a)). However, a.) is included because it is universally true as a well-known  intrinsic property of the vacuum state ([9],[10]) and we wish to preserve it for our class of states.

In the models treated so far, such as the free electromagnetic field  and the conformally invariant scalar field [7] enclosed by parallel plates, the energy-momentum tensor was traceless and, due to this condition, Dirichlet and Neumann b.c. were bona fide b.c. (see, also, ref. [5]). In this typical situation, a.) was true but not c.) - the energy density was also negative - so that b.) was  not realized. In both cases the associated quantum field was massless, and the same happens for a massless scalar quantum field. We were thus led to study the case of a \textbf{massive} quantum scalar field, i.e., of mass $\mu\ne0$, about which only incomplete results (with periodic b.c.) exist [8]. This is also the case for which a cosmological constant emerges  in the analysis of [1]. 

Our result is that a. and b.) (and consequently c.)) are true for a state consisting of the vacuum and some zero momentum excitations, if the size $L$ of the cube is of the order of the Compton wave-length of the particle $1/\mu$. This is reasonable because the vacuum and the particles of zero momentum correspond to lowest energy and $p\le0$. On the other hand a.) and b.) represent very strong constraints: they prevent, in particular, that $L\to 0$, i.e., that the neighborhood $N$ above shrinks to a point.
  
In the next section II we review our definition of the correct energy density operator, which is the basis of all our results. 

In section III we derive the energy density and pressure in closed form for the vacuum state. The result for the pressure involves some remarkable cancellations and is new. It implies property a.).

In section IV we prove properties b.) and c.) for a special state consisting of the vacuum and some excitations (particles) of zero momentum.

Section V is reserved to the conclusion.

It should be stressed that there exists an immense literature on the Casimir effect. In particular, zeta function methods have been studied during several decades and are also, in principle, applicable to the problem treated in section III, see [9], [10] and references given there, in particular, reference [10] for rigorous results. 

Applications of the Casimir effect to Cosmology from a different point of view were made in references [19], [20], [21] and [22]. We refer, in particular, to the very nice review of Elizalde [20] and references given there on the dynamical Casimir effect. See also refs. [23] and [24].

\section{The conceptual framework}

We review in this section the conceptual framework , following [5], for a rigorous, regularization- independent local quantum field theoretic (lqft) treatment of the internal Casimir energy and pressure of a cube for a quantum scalar field of mass $\mu\ne0$.
 
We shall define a \textbf{state} of the field, as is common in lqft [11], as a positive normalized linear functional
$\omega(\cdots)$ such that $\omega(\mathbf{ identity })=1$, $\omega(A^{\dag}A)\ge0$, where $A$ a bounded observable, and 
$A^{\dag}$ its adjoint. Occasionally, $A$ will be taken as an unbounded operator, assuming that $\omega(A)$ is well defined, which will be shown to be the case when $A$ is the (renormalized) Hamiltonian density and $\omega$ a special state. Examples of states are vacuum states and temperature states [9], or 
$$
\omega_F(A)=(\Omega_F,A\Omega_F)
\eqno{(2)}
$$
the Fock vacuum state, where $\Omega_F$ is the Fock vacuum.
Let $\Phi(x)$ denote a scalar quantum field of mass $\mu\ne0$ ($\hbar=c=1$) on symmetric Fock space $\cal{F}$:
\begin{eqnarray*}
\Phi(x)=\frac{1}{(2\pi)^{3/2}}\int\,d^3k\frac{1}{\sqrt{2\omega_k}}[a_{\vec{k}}e^{-ik.x}+a^{\dag}_{\vec{k}}e^{ik.x}]
\end{eqnarray*}
$$\eqno{(3)}$$
where $x=(x_0,\vec{x})$, $k=(k_0,\vec{k})$, $k.x=k_0x_0-\vec{k}\cdot\vec{x}$, $a,a^{\dag}$ are the usual annihilation and creation operators and
$$
k_0=\omega_k=\sqrt{k^2+\mu^2}
\eqno{(4)}
$$
with $k=\vert \vec{k} \vert $. Occasionally, we shall use a cutoff version of (1),
$$
\Phi_\Lambda(x)=\frac{1}{(2\pi)^{3/2}}\int\,d^3k\frac{1}{2\omega_k}[a_{\vec{k}}e^{-ik.x}+a_{\vec{k}}e^{ik.x}]C_\Lambda(k)
\eqno{(5)}
$$
where $\Lambda^{-1}\ge0$ is a cutoff with dimension of length, and $C_\Lambda$ is a smooth function whose properties will be specified later. The associated formal Hamiltonian density is
$$
T_{00}(x)=\frac{1}{2}\lbrack (\partial\phi(x)\partial x_0)^2+(\nabla\phi(x))^2+\mu^2(\phi(x))^2\rbrack
\eqno{(6)}
$$
We shall be interested in the expectation value of $T_{00}(x)$ in certain states. Taking as a state the Fock vacuum state (1), we find for the expectation value of the formal Hamiltonian density
$$
\omega_{F,00}(f)=\int\,d\vec{x}\omega_F(T_{00}(x_0,\vec{x})f(\vec{x})=\,
=\frac{1}{2}\int\,d^3k\tilde{f}(\vec{k})\omega_k
\eqno{(7)}
$$
where $\omega_k$ is given by (3) and $\tilde{f}$ denotes the Fourier transform of $f$. The necessity of using a test function f - as usual in lqft, 
$f\in\cal{S}$, where $\cal{S}$ is the Schwartz space of infinitely differentiable functions of fast decrease [11]- is due to the fact that $\omega_{F,00}$ is, for fixed time $x_0$, a distribution, an essential feature of lqft, and the limit
$f\to\delta$, $\delta$ being the delta function, is ill-defined: there is a well-known reason for that, both mathematical and physical, see ([12], p.35) for a nice elementary physical discussion. Unfortunately, there is a too large indeterminacy associated to the test function $f$ in (7). This is one of the basic problems in applications to Cosmology, where we wish to compute a ``local'' value of the energy density, at least closest possible to the approximation  of perfect homogeneity, which is surprisingly successful [13].

In order to attack the above problem, we attempt to ``restrict'' the quantum field to certain simple compact regions, characterized by some fixed number with the dimensions of length, which we denote by $L$. The simplest choice is a cube of side $L$,with a collection of such cubes providing a cover of Euclidean space. With such a choice, $L$ still remains as a free parameter, but the indeterminacy is widely reduced. What is, however, the correct formula for the Hamiltonian density in this case? On Fock space $\cal{F}$, the quantity $T_{00}(x)$, formally defined by the r.h.s. of (6), is not a well-defined operator-valued distribution: only the Wick-ordered density 
$$
H(x)= :T_{00}(x):=T_{00}(x)-\omega_F(T_{00}(x))
\eqno{(8)}
$$
is one. In going over from (8) to a compact region $K$ with boundary $\partial K$, it is convenient to rewrite the Wick product in (8) using the point-splitting technique ([4],p.316;[6], appendix) and reexpressing the field in terms of different creation and annihilation operators $\tilde{a}_{\vec{k}}$ associated to the eigenfunctions given below in (11d),(11e)(see (10) of [4] for details). These fields, denoted by $\tilde{\phi}$, define a new representation, corresponding to a different observable algebra ( the Hilbert space of test functions is no longer $L^2((\mathbf R)^3)$, see ([5], (2.13) et ff),  a new density 
$\widetilde{T_{00}(x)}$, which results from writing (6) in the new representation, and a new vacuum $\Omega$, annihilated by the new absorption operators associated to the eigenfunctions. The latter defines a new vacuum state $<\cdots>_{vac}$ given
by
$$
<A>_{vac}=(\Omega, A \Omega)
\eqno{(9)}
$$
We write now $\widetilde{T_{00}(x)}$ as a sum of a Wick-ordered term with respect to $\Omega$ - denoted by a semicolon- and the $\Omega$- vacuum energy $(\Omega,\widetilde{T_{00}}\Omega)$, obtaining (this is the precise analogue of formula (13a) of [4]
for the case $\mu\ne0$):
$$
\widetilde{H(x)}=;\widetilde{T_{00}(x)};+(\Omega, \widetilde{T_{00}(x)}\Omega)-\omega_F(T_{00}(x))
\eqno{(10)}
$$
We now obtain from (10) (for details, see, again [3], [4] or [5]):
     $$	\left\langle\widetilde{H(x)}(x)\right\rangle_{vac}=\frac{1}{2}\lim_{y\rightarrow x}\frac{\partial^2}{\partial x_0^2}\left\{D_0^{(+)}(x-y)-D_k^{(+)}(x,y)\right\}
\eqno{(11a)}
$$
where 
$$D_0^{(+)}(x)\equiv(2\pi)^{-3}\int{d\vec{k}\;e^{-i\left(k_0x_0-\vec{k}\cdot\vec{x}\right)}}
\eqno{(11b)}
$$
$$
	D_k^{(+)}(x,y)=D_k^{(+)}(x_0\!-y_0,\vec{x},\vec{y})=i\sum_{k_n\!\in A}\frac{1}{2\omega_{k_n}}u_{k_n}(\vec{x})\overline{u}_{k_n}(\vec{y})\cdot e^{-i\omega_n\left(x_0\!-y_0\right)}
\eqno{(11c)}
$$
, with $\left\{u_{k_n}\right\}_{k_n\!\in A}$ denoting a complete set of normalized eigenfunctions of the Laplacian in $K$, satisfying the prescribed b.c. on $\partial K$, i.e.
$$
	-\Delta u_{k_n}(\vec{x})=\omega_{k_n}^2u_{k_n}(\vec{x})
\eqno{(11d)}$$
Above, $A$ is a discrete set which depends on the chosen b.c., and
$$
	\omega_{k_n}=\left(k_n^2+\mu^2\right)^{1/2}
\eqno{(11e)}$$

It should be remarked that the distribution $\omega_{F}(T_{00}(x))$ in (8) is preserved in (10): this fact is very important and will be referred to as \textbf{normalization condition} [3]: it is reflected by the presence of the distribution term
$D_0^{+}$ on the right hand side of (11a). It is the only possible universal normalization condition, because only the field $\phi$ in infinite (Minkowski) space-time is a standard (universal) object. This is the justification of (10), (11a), on
which all our results are based.

For a conformally invariant scalar quantum field, which is traceless, i.e.

$$
	g^{\mu\nu}T_{\mu\nu}=0
\eqno{(12a)}
$$
the r.h.s. of (11a) is finite and homogeneous for Dirichlet (and Neumann) b.c. [8]. This is due to a remarkable cancellation brought about by (12a)[8]. In applications to Cosmology, 

$$
	\left\langle T_{\mu\nu}\right\rangle_{vac}=g_{\mu\nu}\rho_{vac}=diag\left(\rho_{vac},p_{vac},p_{vac},p_{vac}\right)
\eqno{(12b)}
$$
with $g_{\mu\nu}=diag(1,-1,-1,-1)$, and thus (12a) is not valid: see section IV. For this reason we are led to study an (ordinary) quantum scalar field, for which Dirichlet and Neumann b.c. are incompatible with the general principles of lqft [5]. This fact, not generally appreciated, was also emphasized by Hagen [14] with a different wording, i.e., that there is no justification to eliminate the ensuing divergent surface terms. These result from the wild fluctuations of quantum fields over sharp surfaces [15].

There remains the choice of periodic b.c. as the only sufficiently ``soft'' b.c., which we adopt as in [5]. This choice is technical, but has a nice property in the applications to Cosmology: the ensuing energy density may be shown to be uniform 
i.e, independent of $x$, and equals the average value of the energy in a cube of side $L$ containing the point, with periodic b.c.. The parameter $L$ plays the role of the test function in (7) and is to be chosen as small as the local structure of quantum fields permits, approaching as closely as possible the idealized condition of perfect homogeneity of classical Cosmology.

In order to find upper and lower bounds on $L$ we need, however, to compute the r.h.s. of (11a). For that purpose we introduce the cutoff field (5) and prove in the forthcoming section III that, for a large class of cutoff functions $C_\Lambda$, we obtain a uniform energy density (25) and the thermodynamic pressure (27) in closed form.

\section{The internal Casimir effect for the massive scalar field in a cube}

We consider a cube  of side $L$ adopting periodic b.c., so that the set $A$ in (11c) is 
 $A=\left\{\frac{2\pi}{L}\vec{n},\;\vec{n}\in\mathbb{Z}^3\right\}$, with eigenfunctions in (11d), given by
$$
		u_{\vec{n}}(\vec{x})=L^{-3/2}\,e^{\left(2i\frac{in_1\pi}{L}x+2i\frac{n_2\pi}{L}y+2i\frac{n_3\pi}{L}z\right)}
\eqno{(13a)}
$$
and $\omega_{k_n}$ in (11e) of the form
$$
	\omega_{\vec{n}}=\left[\left(\frac{2\pi}{L}\right)^2\vec{n}^2+\mu^2\right]^{1/2}
\eqno{(13b)}
$$
In order to evaluate the r.h.s. of (10) and (11a), we use regularized fields (5) as well as regularized fields 
$\tilde{\phi_\Lambda}$ corresponding to the fields $\tilde{\phi}$ in the new representation. Let 
$<T_{00}(x)>_{vac,L,\Lambda}$ denote the r.h.s. of (11a) but with $D_0^{+}$ (resp. $D_k^{+}$) replaced by the two-point functions of the regularized fields $\phi_\Lambda$ (resp. $\tilde{\phi_\Lambda}$), and 
$$
C_{\Lambda}(x)=C(\Lambda \times x)
\eqno{(14a)}
$$ 
We now have

\textbf{Theorem1}

Let $C$ in (14a) be a real-valued smooth function satisfying the conditions:
$$
C(0)=1
\eqno{(14b)}
$$
$$
\int_0^{\infty} C^{(k)} dx < \infty \forall k=1,2,\cdots
\eqno{(14c)}
$$
$$
\lim_{x\to\infty}C^{(k)}(x)=0 \forall k=1,2,\cdots
\eqno{(14d)}
$$
Then
\begin{eqnarray*}
\rho_{vac,L}&=&\lim_{\Lambda\to0}\left\langle T_{00}(x) \right\rangle_{vac,L,\Lambda}\\
           &=&-L^{-4}\left\{\frac{2r}{(2\pi)^2}\sideset{}{'}\sum_{\vec{m}\in\mathbb{Z}^3}{m^{-3}K_1(rm)}+\frac{r^2}{2}\sideset{}{'}\sum_{\vec{m}\in\mathbb{Z}^3}{m^{-2}\left[K_0(rm)+K_2(rm)\right]}\right\}
\end{eqnarray*}
$$\eqno{(14e)}$$

where
$$
	r=L\mu
\eqno{(14f)}
$$

independently of $C_\Lambda$ satisfying (14a-d).

The proof of Theorem1 is identical to the proof of regularization independence of the ``finite part'' of the energy - which is identical to the whole energy in the case of periodic b.c., see ([5], Remark 3.6) - for $\mu=0$ in ([5], Theorems 3.1 and 3.2) . That proof is an adaptation of the remarkable treatment in G.H. Hardy's book [16]. We therefore omit the details and derive the results adopting a special convenient regularizer:
$$
C(x)=e^{-x/2},x\ge0
\eqno{(15)}
$$   
 obtaining for the corresponding cutoff version of the r.h.s. of (11a):
$$
	\rho_{vac,\Lambda}\equiv\left\langle\widetilde{H(x)}(x)\right\rangle_{vac,L,\Lambda}= \frac{1}{2}\left\{-\frac{1}{(2\pi)^3}\int{d\vec{k}\,\omega_{\vec{k}}\,e^{-\Lambda\omega_{\vec{k}}}}+I_{cube}(\Lambda)\right\}
\eqno{(16)}
$$
where
$$
	I_{cube}(\Lambda)=L^{-3}\sum_{\vec{n}}{\omega_{\vec{n}}\,e^{-\Lambda\omega_{\vec{n}}}}
\eqno{(17)}
$$
and
$$
\omega_{\vec{k}}=\left[\vec{k}^2+\mu^2\right]^{1/2}
\eqno{(18)}
$$
Using, now, the Poisson summation formula on (17):

$$
\sum_{\vec{n}\in\mathbb{Z}^3}{f(2\pi\vec{n})}=(2\pi)^{-3/2}\sum_{\vec{m}\in\mathbb{Z}^3}{\hat{f}(\vec{m})}
\eqno{(19a)}
$$

where

$$
\hat{f}(\vec{m})\equiv (2\pi)^{-3/2}\int{d\vec{k}\,e^{-i\vec{m}\cdot\vec{k}} f(\vec{k})}
\eqno{(19b)}
$$

with

$$
f(\vec{k})=\left[(\vec{k}/L)^2+\mu^2\right]^{1/2}e^{-\Lambda\left[\left(\frac{\vec{k}}{L}\right)^2+\mu^2\right]}
\eqno{(19c)}
$$

\vspace{0.05cm}

Changing from cartesian to spherical coordinates, one obtains from the terms with $m\neq0$ in (19b):
\begin{eqnarray*}
\hat{f}(\vec{m}\!\neq\!\vec{0})&=&(1/(2\pi^2))\int_{0}^{\infty}{dk\,k^2\left(\int_{-1}^1{e^{-imku}du}\right)f(k)}\\
	&=&(1/(im(2\pi)^2))\int_0^{\infty}{dk\,kf(k)\left(e^{imk}-e^{-imk}\right)}
\end{eqnarray*}
$$\eqno{(20a)}$$
where $m\equiv|\vec{m}|$. From the term $m=0$ in (19b)
\begin{eqnarray*}
	\hat{f}(\vec{m}\!=\!\vec{0})&=&\frac{4\pi}{(2\pi)^3}\int_0^{\infty}{dk\,k^2f(k)}
\\
	&=&\frac{4\pi}{(2\pi)^3}\int_0^{\infty}{dk\,k^2\left((k/L)^2+\mu^2\right)^{1/2}e^{-\Lambda\left((k/L)^2+\mu^2\right)^{1/2}}},\\
	&=&\frac{L^3}{(2\pi)^3}\int{d\vec{k}\,\omega_{\vec{k}}\,e^{-\Lambda\omega_{\vec{k}}}}
\end{eqnarray*}
$$\eqno{(20b)}$$
One sees that the term $m=0$ (20b) is exactly the first term in (16) but with opposite sign. Thus,

\begin{eqnarray*}
     \rho_{vac,\Lambda}&=&\frac{1}{2}L^{-3}\sum_{\vec{n}}{\omega_{\vec{n}}\,e^{-\Lambda\omega_{\vec{n}}}}
\\
	&=&\frac{1}{2}L^{-3}\sideset{}{'}\sum_{\vec{m}\in\mathbb{Z}^3}\frac{1}{im(2\pi)^2} \int_0^{\infty}{dk\,k\left((k/L)^2+\mu^2\right)^{1/2}e^{-\Lambda\left((k/L)^2+\mu^2\right)}\left(e^{imk}-e^{-imk}\right)}
\end{eqnarray*}

Above $\sum^{'}$ denotes exclusion of the vector $\vec{m}=\vec{0}$. Making an obvious change of variable on the above, we arrive at

$$
\rho_{vac,\Lambda}=\\
=\sum_{\vec{m}\in\mathbb{Z}^3}\frac{2L^{-2}}{m(2\pi)^2}\frac{\partial}{\partial m}\frac{\partial^2}{\partial\Lambda^2} \left[\int_0^{\infty}{dk\left(k^2+\mu^2\right)^{-1/2}e^{-\Lambda\left(k^2+\mu^2\right)^{1/2}}\cos(mLk)}\right]
\eqno{(21)}
$$

It is known [17]  that
$$
g(y)=\int_0^\infty{dx\,cos(xy)(x^2+\alpha^2)^{-1/2}e^{-\beta\sqrt{x^2+\alpha^2}}}=K_0\left[\alpha(\beta^2+y^2)^{1/2}\right]
\eqno{(22)}
$$

where $K_{\nu}$ ($\nu=0,1,2\ldots$) denote the modified Bessel functions [17]. Introducing (22) into (21), we obtain

$$
\rho_{vac,\Lambda}=-\sideset{}{'}\sum_{\vec{m}\in\mathbb{Z}^3}\frac{2L^{-2}}{m(2\pi)^2}\frac{\partial}{\partial m}\frac{\partial^2}{\partial\Lambda^2} \left\{K_0\left[\mu\left(\Lambda^2+m^2L^2\right)\right]\right\}
\eqno{(23)}
$$

One may easily calculate the derivatives in (23) and take the limit $\Lambda\to0_+$ of $\rho_{vac,\Lambda}$ to get:

$$
\rho_{vac,L}\equiv\left\langle T_{00}(x)\right\rangle_{vac,L}=-\frac{2L^{-2}}{(2\pi)^2}\sideset{}{'}\sum_{\vec{m}\in\mathbb{Z}^3}\left\{\mu m^{-3}L^{-1}K_1(\mu mL)+ \frac{\mu^2m^{-2}}{2}\left[K_0(\mu mL)+K_2(\mu mL)\right]\right\}
\eqno{(24)}
$$

\vspace{0.2cm}

Notice that, even for periodic b.c., if the normalization condition (term $D_0^{+}$ in (11a)) is not imposed, there is no cancellation of the term (20b) and the result diverges as $\Lambda^{-4}$. This yields a result which is off by a factor of order $10^{55}$ upon taking $\Lambda$ equal to the inverse of the Planck length [25].

This proves (14e) and hence theorem 1 for the special regularizer (15). As remarked before, the proof of regularization independence is omitted.

\vspace{0.2cm}
Let

$$
E_{vac,L}=L^3 \rho_{vac,L}
\eqno{(25)}
$$

denote the energy in a cube of side $L$.

For the thermodynamic pressure (since $V=L^3$):
\begin{eqnarray*}
	p_{vac,L}&=&-\frac{\partial E_{vac,L}}{\partial V}=-\frac{\partial E_{vac,L}}{\partial L}\frac{\partial L}{\partial V}=-\frac{1}{3L^2}\frac{\partial E_{vac,L}}{\partial L}
\\
	&=&\frac{2}{3L^2(2\pi)^2}\sideset{}{'}\sum_{\vec{m}\in\mathbb{Z}^3}\mu^2m^{-2}\left\{K'_1+ \frac{1}{2}\left(K_0+K_2\right)+ \frac{1}{2}\mu mL\left(K'_0+K'_2\right)\right\}
\end{eqnarray*}
$$\eqno{(26)}$$

Since $K'_1=-(K_0+K_2)/2$, \hspace{0.1cm}$K'_2=-(K_1+K_3)/2$ \hspace{0.1cm}and $K'_0=-K_1$, the first and second terms cancel each other (this is the remarkable cancellation mentioned in section II, which leads to the results of section IV):

\begin{eqnarray*}
	p_{vac,L}&=&\frac{1}{3L(2\pi)^2}\sideset{}{'}\sum_{\vec{m}\in\mathbb{Z}^3}\mu^3m^{-1}\left[-K_1(\mu mL)-\frac{K_1+K_3}{2}\right]
\\
	&=&-\frac{1}{3L(2\pi)^2}\sideset{}{'}\sum_{\vec{m}\in\mathbb{Z}^3}\mu^3m^{-1}\left[\frac{3}{2}K_1(\mu mL)+\frac{1}{2}K_3\right]
\end{eqnarray*}

Rewriting $p_{vac,L}$ using the definition (14f),
$$
	p_{vac,L}=-L^{-4}\left\{\frac{r^3}{3(2\pi)^2}\sideset{}{'}\sum_{\vec{m}\in\mathbb{Z}^3}{m^{-1}\left[\frac{3}{2}K_1(rm)+\frac{1}{2}K_3(rm)\right]}\right\}
\eqno{(27)}
$$
\vspace{0.3cm}

\section{Cosmological constant}

As is well-known, it is the advent of quantum field theory which made consideration of the Cosmological constant obligatory, not optional, as discussed by Zel'dovich many years before the acceleration of the expansion of the Universe was discovered [25]. Indeed, the only possible covariant form for the energy of the quantum vacuum is (12b), which is mathematically equivalent to the Cosmological constant $\Lambda$ by

$$
\Lambda= 8 \pi G \rho_{vac}
\eqno{(28)}
$$

(12b) takes the form of a perfect fluid with energy density $\rho_{vac}$ and isotropic pressure $p_{vac}=-\rho_{vac}$ which corresponds to (1b) of the introduction. Note that (1a) is universally true as an intrinsic property of the quantum vacuum
([9],[10]), which, together with (1b), implies (1c).

We now investigate whether (1b) and (1c) hold in the model of section III. The answer is clearly negative:

Since $K_i(x)\geq0$ for all $i$ and $x$ [17], by (25) and (27) both $\rho_{vac,L}$ and $p_{vac,L}$ are negative. Thus (1b) and (1c) are violated whatever $L$.

In order to ensure a property corresponding to (1c) (our (1c')), we propose to add some one-particle excitations of zero momentum to the vacuum energy: they are the (one-particle) excitations of lowest energy compatible with pressure $p\leq0$. The  states of zero momentum  are  compatible with periodic b.c. and have  energy $\mu$; the corresponding total energy $E_L$ is
$$
	E_L=E_{vac,L}+\mu
\eqno{(29a)}
$$
We write
$$
	\tilde{\rho}_{vac}=\tilde{\rho}_{vac,L}=L^{-3}E_L
\eqno{(29b)}
$$
The corresponding state of the quantum field is (where $A$ is any observable):
$$
\left\langle A \right\rangle_{vac,L}^{'}= ( (\tilde{a}_{\vec{0}})^{\dag}\Omega, A (\tilde{a}_{\vec{0}})^{\dag}\Omega )
\eqno{(30)}
$$
and corresponds to one particle of zero momentum in each cube, and thus to a one particle energy density $\mu L^{-3}$.

We now take $\mu$ as a \emph{fixed} given quantity and, assuming (27), inquire into the possible values of $L$. There are three cases :

a.)$r\ll1$

In this case
$$
	L\ll 1/\mu
\eqno{(31a)}
$$
and we may take the zero mass limit in $\rho_{vac,L}$ ($\mu\rightarrow0+$), obtaining from (24),(25) for $E_{vac,L}$:
\begin{eqnarray*}
	E_{vac,L}&=&-\frac{2L}{(2\pi)^2}\sideset{}{'}\sum_{\vec{m}\in\mathbb{Z}^3}\left(L^{-2}m^{-4}+ \frac{2}{2}L^{-2}m^{-4}\right)
\\
	&=&-\frac{4L^{-1}}{(2\pi)^2}\sideset{}{'}\sum_{\vec{m}\in\mathbb{Z}^3}\frac{1}{m^4}=-0.52L^{-1}
\end{eqnarray*}
$$\eqno{(31b)}$$
Above, we used (for $z\rightarrow0+$)
$$
	K_0\sim-\ln{z}
\eqno{(31c)}
$$ 
and
$$
	K_{\nu}\sim\frac{1}{2}\;\Gamma(\nu)\left(\frac{1}{2}z\right)^{-\nu}
\eqno{(31d)}
$$

The condition
$$
	\tilde{\rho}_{vac}>0
\eqno{(1c')}
$$

in (27) (corresponding to (1c)) yields, by (31b),
$$
	\mu\geq0.52L^{-1}
\eqno{(31e)}
$$

in contradiction with (31a). Note that \emph{no} principle of lqft sets a lower limit to the localization region, except the limit $L\rightarrow0$, corresponding to point-like fields: in this case, $\rho_{vac,L}\!\to\!-\infty$, by (25), which agrees with a theorem of Epstein, Glaser and A. Jaffe [6];

b.)$r\gg1$ 

In this case exponential decay of the $K_i(x)$ [17] yields zero energy density and pressure. This agrees with what is expected from Newton-Wigner localization of quantum fields [18];

c.)$r\approx1$

The adequacy of this case could be checked by comparing the results with (1c') and (1b). From (24), (29) one concludes that

$$
	\tilde{\rho}_{vac,L}=-\frac{1}{L^{4}}\left(\frac{1}{(2\pi)^2}\sideset{}{'}\sum_{\vec{m}\in\mathbb{Z}^3}{\left\{2rm^{-3}K_1(rm)+ r^2m^{-2}[K_0(rm)+K_2(rm)]\right\}-r}\right)
\eqno{(32)}
$$

From  Fig.\ref{graficorho} below, one can see that (1c') is satisfied in the interval used. Since the case is $r\approx1$, we used the range $r=[0.5,5]$.
\begin{figure}[ht]
\begin{center}
\includegraphics [height=7 cm]{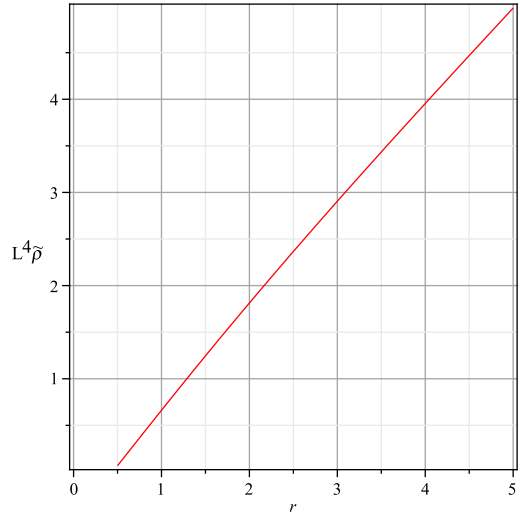}
\end{center}
\caption{Dark Energy density multiplied by $L^4$ as function of $r=\mu L$}
\label{graficorho}
\end{figure}

\begin{figure}[ht]
\begin{center}
\includegraphics [height=7 cm]{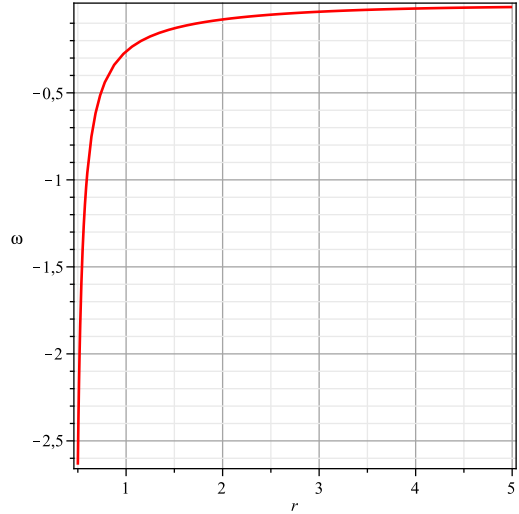}
\end{center}
\caption{$w$ as function of $r=\mu L$}
\label{graficoomega}
\end{figure}

Let us require that
$$
	w=-\frac{p_{vac}}{\tilde{\rho}_L}
\eqno{(33)}
$$
We use (32) and (27) to obtain  Fig.\ref{graficoomega}. It shows that (33) is also satisfied, with
$$
-1\le w<-1/2
\eqno{(34)}
$$
 in a range of $r$ given by  $r\approx[0.592,0.739]$, where $r\approx0.592$ results in a flat (Minkowski) Universe (w=-1).
The range (34) is of relevance in Cosmology because, under certain assumptions, it is sufficient for the existence of an accelerating Universe [29].

Thus, if we assume

(a.)positivity (1c');
 
all values of $r>0.444$ are allowed, according to Fig.1. Assuming, however, in addition:

(b.)the Cosmological constant condition
$$
	p_{vac,L}=-\tilde{\rho}_{vac,L}
\eqno{(35a)}
$$
(i.e.,(33), for $w=1$), we see that there results a \textbf{definite} nonzero value for $\tilde{\rho}_{vac,L}$ given by

$$
r\approx0.592
\eqno{(35b)}
$$

For very large boxes ($L\gg\mu$) the trivial solution satisfying both (a.) and (b.), 
$p_{vac,L}=\tilde{\rho}_{vac,L}=0$ is obtained. 

\textbf{Remark1}

Since the energy density (24) continues to be uniform by any translation (of the origin) or rotation of the coordinate axes,the state obtained from (30) by ergodic means over the parameters of the Euclidean group ($(\vec{a},R)$) yields the same Hamiltonian density (24). There exists, therefore, a Euclidean invariant state of the quantum field in which the expectation value of the energy-momentum tensor is of the form (12b), with (1c') satisfied.

We have therefore shown that the state (30) (or its Euclidean invariant version, see remark) provides a concrete quantum field model of  the Cosmological constant ((1b), (28)) with, in addition, the property (1c').

We now come back to the introduction and discuss the possible significance of our results in the application of the dark energy problem in Cosmology.

Dark energy X is the biggest challenge for the New Cosmology (see [27] for a readable review). Together with the distinguishing feature of emitting no light, it has large negative pressure $p_X$. Data are consistent with
$$
	p_X = w_X\rho_X
\eqno{(36a)}
$$

with

$$
w_X=-1
\eqno{(36b)}
$$

where $\rho_X$ denotes energy density of X, and

$$
	\rho_X\cong2.7\;x\;10^{-47}\;GeV^4
\eqno{(37)}
$$

It is also well-known to be approximately homogeneous . When considering the coupling to gravity, the vacuum energy density $\rho_{vac}$ acts, as we have seen, like a Cosmological constant $\Lambda$ equal to (28), and we have presently constructed a state of the quantum field with this property. How does it fit in the general cosmological theory of dark energy [27], and is it, at least roughly, consistent with the data coming from the analysis of distant type Ia supernovae [31] and e.g. balloon measurements of the anisotropy of the cosmic microwave background [32]?

Under certain conditions [26], the back-reaction of the quantum field on the space-time geometry is described by the ``semiclassical Einstein equation'' [26], which yields for the time-dependent scale factor (see, e.g., [33]):

$$
\ddot{R}(t)=-\frac{4 \pi G}{3}(\rho + 3p)R(t)+\frac{\Lambda}{3}R(t)
\eqno{(38)}
$$

where $\Lambda$ is given by (28), and $p$ is the pressure of all forms of matter. Instead of (28), we replace $\Lambda$ in (38) by

$$
\Lambda=8 \pi G \tilde{\rho}_{vac,L}
\eqno{(28')}
$$

where $\tilde{\rho}_{vac,L}$ is given by (29b), and may be written, by (30),

$$
\tilde{\rho}_{vac,L}= \langle\widetilde{H(x)}\rangle^{'}_{vac,L}
\eqno{(39)}
$$

(10), (11e) and the formal expression

$$
;\tilde{T_{00}(x)}; = \sum_{n} \omega_{k_n} a^{\dagger}_{k_n} a_{k_n}
\eqno{(40)}
$$

yield, together, (29a).

Let us assume that a sea of slowly moving elementary particles left over from the earliest moments - the hypothetical axions ([28], [29]) of mass [28]

$$
\mu_a \approx 10^{-5}- 10^{-4} eV = 10^{-13}- 10^{-14} GeV
\eqno{(41)}
$$
are described by a scalar quantum field, corresponding to an initial state (30), with $L$ given by (26) and (35b), which, by (41), yields the value

$$
L \approx 5.92 \times 10^{12}- 10^{13} GeV^{-1}
\eqno{(42)}
$$

with the estimate for $\rho_X=\rho_{\Lambda}$, here identified with $\tilde{\rho}_{vac,L}$:

$$
\rho_X= 1.44 \times 10^{-56}- 10^{-52} GeV^4
\eqno{(43)}
$$

By (42),

$$
L \ll L_H \approx 10^{42}  GeV^{-1}
\eqno{(44)}
$$

with $L_H$ denoting the Hubble length [13], proving (1a) of the introduction.

Is the Cosmological constant, defined by the r.h.s. of (28'), really a constant, i.e., invariant under time evolution? 
Assuming that the energy equation which follows from Friedmann's equation ([13], [33])
$$
\frac{du}{dR}=-3pR^2
\eqno{(45)}
$$
where $ u= \rho R^3 $ is the total mass, is also valid for the individual cosmic components $X$, we obtain from (36a), (45):
$$
\rho_XR^{3(1+w_X)}=c
\eqno{(46)}
$$
where $c$ denotes a constant, from which follow that, if $w_\Lambda=-1$,
$$
\rho_\Lambda=c
\eqno{(47)}
$$
where $c$ again denotes a constant. It follows that if we choose $L$ as above for the initial state of the quantum field (30), (35a) and $\tilde{\rho}_{vac,L}$, given by (32), are constant throughout evolution, and therefore do define a constant $\Lambda$ on the r.h.s. of (38), by (32) and (28'). The global topology of the Universe is, of course, determined by (38), and has nothing to do with the periodic b.c. on the state, which, in the present model, does not change with (cosmic) time and was just used to determine $\Lambda$. This is consistent with the fact that $L$ only depends on the mass of the particle and, by (42), lies in the submillimeter range: it is the smallest possible dimension compatible with both the local structure of the underlying quantum field and conditions (1a)-(1c). It is entirely negligible in comparison with the Hubble length, and therefore the assumption of infinite periodicity is harmless.

We close this section with some remarks establishing a link to the introduction: the density computed on the state (30) given by (39) corresponding to any point $x$ is the energy density of any cube containing the point $x$ with periodic b.c.. The pressure is the thermodynamic pressure corresponding to this energy, defined in (26), corresponding to what an observer at $x$ would measure in a Gedankenexperiment. From this point of view, the partition into cubes may be dispensed with: it served as intermediary step to construct a state of the quantum field depending on a parameter $L$ with the stated properties. In particular, the energy density and the pressure do not depend either on the time or on the space variable, only on $L$, and therefore are homogeneous, as expected for a theory of dark energy in the approximation of perfect homogeneity (see, e.g., [13]), which has, of course, nothing to do with the real physical cosmological homogeneity, whose scale is of order $70-100 H^1$ [30], and is traditionally accounted for theoretically by cosmological perturbation theory (see, e.g., [33]).

\section{Conclusion}

In this paper we used the conceptual framework introduced in [5] to obtain closed form expressions for the energy density and the pressure of a massive quantum scalar field with periodic  b.c. on a cube, which completes the analysis of [8]. As an application, we showed that positivity (1c') and the cosmological constant property (1b) hold true for a special state of the quantum field consisting of the vacuum and (a certain density of) one-particle excitations of zero momentum, if the length of the cube is a fixed number of the order of the Compton wavelength of the particle.

We consider two points of special relevance in our work. The first one was to point out how strong the constraints of negative pressure and the cosmological constant condition are: they are, in particular, inconsistent with the catastrophic limit $L\to 0$, because they require positivity of the energy density (1c). The second basic point is the crucial role played by the condition of nonzero mass $\mu\ne0$: for zero mass, if one attempts to make $\tilde{\rho}$ positive by adjoining low energy one-particle excitations, the resulting pressure will be positive and (1c') can never be realized. In this connection, it is of interest to remark that the emergence of an effective cosmological constant in [1] seems to require positive mass. 

Our tentative approach to dark energy hinges on the existence of a hypothetical scalar particle. Even in the hypothesis of
detection of the axion (discussed in detail in ([29], section 10.5.1), several difficult problems would remain open, in particular, what would be the reason for the extraordinary stability of the hypothetical initial state (30)? In this respect,
the analysis of [1], based on Hadamard states, goes much deeper: it is also to be stressed that we assumed the existence of a cosmological constant, and looked at the compatibility of this assumption with a state of the massive quantum scalar field (see also [23],[24]), while the authors of [1] obtained it from first principles. On the other hand, the analysis of [1] seems to contain several a priori undetermined constants, while ours is able to make a definite prediction on the range of the particle masses as to yield a good comparison with experimental data. In fact, the numbers of [28] yield a good comparison with the latter if the state (30) is assumed, as shown at the end of section IV.

As a technical open problem, we observe that Theorem 1 does not prove that the distribution at the r.h.s. of (11a) is the function (14e), although this is to be expected.

\section{Acknowledgment}
We are grateful to the referee for remarks which led to an important conceptual reformulation of a previous version of the paper.

\end{document}